\documentclass[lettersize,journal]{IEEEtran}
\usepackage{amsmath,amsfonts}
\usepackage{algorithmic}
\usepackage{array}
\usepackage[caption=false,font=normalsize,labelfont=sf,textfont=sf]{subfig}
\usepackage{textcomp}
\usepackage{stfloats}
\usepackage{url}
\usepackage{verbatim}
\usepackage{graphicx}
\def\BibTeX{{\rm B\kern-.05em{\sc i\kern-.025em b}\kern-.08em
    T\kern-.1667em\lower.7ex\hbox{E}\kern-.125emX}}
\usepackage{balance}
\newcommand{\method}{\textbf{STEM3}}
\newcommand{\methodexp}{\textbf{Speech-Text EEG Match-Mismatch Model}}
\begin{document}

\title{Uncovering the role of semantic and acoustic cues in normal and dichotic listening}

\author{Sai Samrat Kankanala,~\IEEEmembership{Student Member,~IEEE,} Akshara Soman, and Sriram Ganapathy~\IEEEmembership{Senior Member,~IEEE}
\thanks{The authors are with the learning and extraction and acoustic patterns (LEAP) laboratory, Electrical Engineering, Indian Institute of Science, Bangalore, India, 560012. This work was performed with grants received from Ministry of Education, India. }
\thanks{Manuscript received June xx, 2025.}}

\markboth{Journal of \LaTeX\ Class Files,~Vol.~14, No.~8, June~2025}%
{Shell \MakeLowercase{\textit{et al.}}: A Sample Article Using IEEEtran.cls for IEEE Journals}

\maketitle

\begin{abstract}
Speech comprehension is an involuntary  task for the healthy human brain, yet the understanding of the mechanisms underlying this brain functionality remains obscure. In this paper, we aim to quantify the role of acoustic and semantic information streams in complex listening conditions. We propose a paradigm to understand the encoding of the speech cues in electroencephalogram (EEG) data, by designing a  match-mismatch (MM) classification task. 
The MM task involves identifying whether the stimulus (speech) and response (EEG) correspond to each other. 
We build a multimodal deep-learning based sequence model (\method{} - \methodexp), which is input with  acoustic stimulus (speech envelope), semantic stimulus (textual representations of speech), and the neural response (EEG data). 
 We perform extensive experiments on two separate conditions, i) natural passive listening and, ii) a dichotic listening requiring auditory attention. Using the MM task as the analysis framework, we observe that - a) speech perception is fragmented based on word boundaries, b) acoustic and semantic cues offer similar levels of MM task performance in natural listening conditions, and c) semantic cues offer  significantly improved MM classification over acoustic cues in  dichotic listening task. The comparison of the \method{} with previously proposed MM models shows significant performance improvements for the proposed approach. The analysis and understanding from this study allows the quantification of the roles played by acoustic and semantic cues  in diverse listening tasks and in providing further evidences of right-ear advantage in dichotic listening.
\end{abstract}


\section{Introduction}
\IEEEPARstart{H}{uman} speech perception is a complex cognitive process involving the integration of  acoustic and contextual cues.
The speech signal carries information in the low-level acoustic descriptors of the signal like the envelope, energy, pitch, modulation rate, intonation and stress, as well as in high-level semantic descriptors of the spoken content, speaker characteristics and environment/ambiance. The speech envelope is the most commonly used acoustic feature for analyzing neural entrainment and is shown to be synchronized with neural oscillations in the auditory cortex \cite{di2015low}. Furthermore, recent research  \cite{poeppel2020speech} suggests that neural entrainment to speech is not limited to the speech envelope as it also extends to other acoustic features such as phonemic content. However, a systematic understanding of the speech attributes that are encoded in the brain is still an open ended question.

In this paper, we investigate the relative role of semantics and acoustics in parsing the input speech under normal and dichotic listening conditions. 
Continuous natural speech stimuli evoke a  brain response that can be detected with neuro-imaging techniques like electroencephalography (EEG) or magnetoencephalography (MEG). The EEG is a non-invasive neural imaging technique that measures electrical activity in the brain by placing electrodes on the scalp \cite{sanei2013eeg}. Multiple studies have demonstrated that EEG, recorded during speech listening tasks, contains information about the speech stimuli \cite{di2015low,broderick2018electrophysiological,soman2019eeg}. In the context of encoding higher-level information, Liberto et al. \cite{di2015low} demonstrated that EEG signals elicit categorical processing of phonemes in continuous speech. 
Additionally, Broderick et al. \cite{broderick2018electrophysiological} showed that EEG captures linguistic representations such as semantic dissimilarity. 
Similarly, Soman et al. \cite{soman2019eeg} reported that the EEG recorded during the listening state carries information about the language of the stimulus.

The design of studies intended to decode the role of different features during speech perception has been an important research area in auditory neuroscience. Early studies \cite{ding2012neural,crosse2016multivariate,wong2018comparison,vanthornhout2018speech} in this area explored linear models to establish a relationship between continuous natural speech and EEG responses, classified as forward, backward, or hybrid models. These models predict either the EEG from speech stimuli or reconstruct the speech from EEG responses. The correlation between the predicted and the reference signals was used as a measure of neural tracking in numerous studies \cite{de2018decoding}. However, due to the non-linear nature of the auditory system \cite{faure2001there}, linear models may be incapable of capturing the stimulus-response interactions during speech perception.
Deep neural networks have therefore been explored to compare and analyze EEG responses for spoken stimuli  \cite{monesi2020lstm,katthi2021deep,de2021auditory,accou2020modeling}. 

The forward/backward models predict one component (stimulus/response) given the other component (response/stimulus). However, the response (EEG) is not fully determined by the input spoken stimulus as the brain responses also encode other body functions. Hence, most of the successful forward-backward models only predict  $20$-$30$\% of the data variance. On the other hand, 
the match-mismatch (MM) task is a compelling task that can uncover the underlying neural mechanisms  \cite{de2021auditory,puffay2022relating}.  In the MM task, the goal is to determine if a segment of brain response (like EEG) matches with a segment of the spoken stimulus that evoked that response.
Given the binary nature of the MM task, the design of models is somewhat easier compared to the tedious regression tasks undertaken by forward/backward models.
The performance in MM task, measured by MM classification accuracy, provides a simpler and more interpretable metric compared to the traditionally employed correlation metrics. 
Further, the MM task accuracy achieved by the machine-learning models~\cite{wang2024self} is high, as even a partial encoding of the stimulus enables the models to predict the MM score. 

In addition to analyzing speech perception during natural speech listening, which is typically explored in MM tasks, this paper  also delves into the analysis of how the brain perceives speech in complex listening conditions, such as in dichotic listening. Auditory attention decoding refers to the process of decoding or identifying the specific auditory stimulus or sound that an individual is paying attention to when presented with a mixture of auditory stimuli. Such auditory attention decoding (AAD) experiments have been used to investigate auditory processing and attention mechanisms in the brain \cite{ding2012neural,mirkovic2015decoding}. One such listening condition is dichotic listening, in which participants are presented with different auditory stimuli simultaneously in each ear. 
This capability of the human brain is termed "The Cocktail party effect", which refers to the ability of the human brain to selectively focus attention on a particular auditory stimulus, while filtering out other competing sounds in a noisy environment.
Our study aims to identify the specific cues to which the subject directs their attention during this task. Additionally, we aim to analyze the influence of word boundary information on auditory attention decoding (AAD).

\subsection{Related works}

In earlier studies employing the mismatch (MM) task, auditory stimuli of a fixed duration ($5$ seconds) were typically processed using a combination of convolutional and recurrent neural network layers, as demonstrated in works such as \cite{monesi2020lstm,puffay2022relating,accou2020modeling}. More recent research has advanced this approach by leveraging distinct EEG responses -- Thornton et al.~\cite{thornton2024decoding} utilized speech-evoked frequency-following responses, while Thornton et al.~\cite{thornton2024detecting} focused on high-frequency gamma band responses, both achieving notable performance improvements. The role of self-supervised speech representations and contextual text embeddings in the MM task was explored by Wang et al.~\cite{wang2024self}, highlighting their significance in enhancing model performance. Meanwhile, Borsdorf et al.~\cite{borsdorf2023multi} demonstrated that multi-head attention mechanisms offer superior classification accuracy for this task. In a distinct contribution, Soman et al.~\cite{akshara2023IS} proposed a deep learning model to investigate stimulus-response alignment at the sentence level during natural speech listening. Their study underscored the importance of incorporating word-level segmentation information. 

 In dichotic listening, prior works \cite{aydelott2015semantic} had analysed the semantic processing and more recent works \cite{elyamany2024top,mayorova2024sex} focused on interhemispheric connectivity and sex-related differences in selective attention. However, our work analyzes dichotic speech perception by posing it as an MM task, where the match data is constructed with the stimulus-EEG pair that the subject was attending to, while the mismatch data is constructed with the unattended pair. 
 In this manner, our proposed \method{} framework models the normal and dichotic listening in a unified framework, allowing us to contrast the  roles played by acoustic and semantic cues in these two conditions. 




\subsection{Scope and Contributions}
In this study, we propose the MM framework for auditory stimulus-response (EEG) modeling  with a sentence-level model that incorporates word segmentation information. 
This framework is designed to capture the neural responses in both mono-aural and dichotic listening scenarios. 
To conduct this study, we utilize three distinct speech-EEG datasets \cite{broderick2018electrophysiological}: two of them collected during a natural speech listening task and the other during a dichotic listening task. 

Our approach employs a deep learning network for stimulus-response modeling, which consists of three separate sub-networks: one for processing the EEG signals, one for processing the acoustic features of the stimulus and the other for processing the semantic features of the stimulus. The EEG sub-network comprises a series of convolutional layers, followed by word boundary-based average pooling, and  a recurrent/transformer layer that incorporates inter-word context.
The proposed model is termed \methodexp{} (\method{}).


The major contributions of this work are:

\begin{itemize}
\item Formulate a paradigm for auditory stimulus-response (EEG) modeling through the MM task in natural listening as well as in dichotic listening, and provide a unified framework to understand the relative role of acoustic and semantic cues that are encoded in the EEG data.
\item In the mono-aural natural speech listening task, the MM task performance of audio signals is enriched by the induction of word boundaries. Further, the textual information provides statistically similar MM performance compared to the acoustic features.
\item In the dichotic speech listening task, the MM task performance of textual features is significantly higher than those from the audio signal, indicating that EEG signals encode higher-level semantic information.
 
\item For both these tasks,  the performance of the multimodal \method{} significantly improves over the speech-EEG models, indicating that the EEG signal jointly encodes the semantic and acoustic contents of the stimulus. 
The comparison with prior works \cite{borsdorf2023multi,wang2024self} also illustrates that the \method{} significantly outperforms these models.  

\item For dichotic listening, we analyze the brain regions that show improved MM task performance, and this analysis confirms the evidences of right ear advantage \cite{geffen1978development} \cite{tanaka2021neurophysiological}. 
\end{itemize}

\subsection{Impact of the proposed study}
The results of these experiments find applications in robotics \cite{okuno2001human,nguyen2016selection}, brain-computer interface (BCI), and hearing aids \cite{lunner2013hearing}. Hearing loss and the associated use of hearing aids are on the rise in the general population \cite{wilson2017global}. The performance of hearing aids is significantly impacted in situations where the background noise involves multiple talkers \cite{zeng2008cochlear}, known as the cocktail party problem.
Addressing the challenge of assisted listening in multi-talker environments could have broad societal benefits in terms of communication and mental health \cite{andrade2018silent}. AAD leverages knowledge of the listener’s auditory intent (attention) to isolate and enhance the desired audio stream, while suppressing others 
\cite{das2017eeg,van2016eeg,geirnaert2020neuro}.

The auditory attention detection algorithms could be integrated with passive Brain-Computer Interface (\textit{p}BCI) systems for applications in education and interactive music performances \cite{belo2021eeg}. An early attempt in this direction was made by Cho et al. \cite{cho2002attention}, who developed an attention enhancement system for ADHD children using EEG biofeedback in a virtual classroom environment. The AAD method suggested in this work can be used to continuously track the attended auditory source and provide feedback to the user \cite{belo2021eeg}. 

\section{Methods} 
\label{sec:methods}
\subsection{Dataset}
\label{sec:Dataset}
We experiment with two publicly available speech-EEG datasets\footnote{https://doi.org/10.5061/dryad.070jc}, released by Broderick et al. \cite{broderick2018electrophysiological}.
These scenarios include subjects listening to uninterrupted, natural speech and a situation where the subjects participate in a dichotic listening task. In the dichotic case, the subjects listen to two distinct audio streams presented simultaneously, one in each of the two ears. 
For the subsequent discussion, we refer to the first dataset as the natural speech dataset (small) and the second one as the dichotic dataset.
In both experiments, the stimuli were presented using Sennheiser HD650 headphones  with the software provided by Neurobehavioral Systems (\url{http://www.neurobs.com}). The participants were instructed to keep their gaze fixed on a crosshair displayed on the screen throughout each trial and to minimize activities such as eye blinking and other motor movements.

To further evaluate and validate our model’s performance in natural settings, we  experimented with the larger publicly available dataset, ``SparrKULee" \cite{accou2023sparrkulee}. SparrKULee consists of natural speech recordings from a larger pool of subjects and in a non-English setting (Native speakers listening to \textit{Flemish} language recordings). 
This expanded subject pool enables us to test on subjects, who are not part of the training dataset. 
We use the training, validation and test splits following the definitions given in \cite{accou2023sparrkulee} and refer to these experiments as natural speech dataset (large). 

\subsubsection{Natural Dataset - Small }
The natural speech dataset small contains electroencephalographic (EEG) data recorded from $19$ subjects as they listened to continuous speech. The subjects listened to a professional audio-book narration by a single male speaker of a well-known work of fiction. 
For each subject, the data consists of $20$ trials of similar length, each containing $180$s of audio. The trials preserved the chronology of the storyline without repetitions or breaks.  The sentence start/end time and the word-level transcription  of the  speech recordings are provided. 
The word segmentation is obtained using a speech recognition-based aligner \cite{gorman2011prosodylab}. 
The EEG data were acquired using $128$-channel BioSemi system, sampled at  $512$Hz, while the audio data is sampled at $16$kHz. Overall, the speech-EEG data amounted to a duration of $19$ hours. 

\subsubsection{Dichotic Dataset }
The dichotic dataset used in this study comprises EEG recordings obtained from $33$ subjects. Each participant listened to a total of $30$ trials, with each trial lasting $60$ seconds. They were presented with two well-known fictional works, where one story was played to the left ear and the other to the right ear. Different male speakers articulated each story. The participants were divided into groups of $17$ and $16$ individuals (with one subject excluded). Each group of listeners was instructed to focus their attention solely on the story presented in either the left or right ear for all the $30$ trials. Following each trial, the participants were required to answer multiple-choice questions about the stories played in left/right ear. To maintain consistency, the audio streams of each story within a trial were normalized to have the same root mean squared (RMS) intensity. To prevent the unattended story from capturing the participants' attention during silent periods in the attended story, any silent gap exceeding $0.5$ seconds was truncated. 

\subsubsection{Natural Dataset - Large }

The SparrKULee  dataset  \cite{accou2023sparrkulee} is a speech-evoked auditory repository of EEG, comprising $64$ channel EEG recordings from $85$ adult ($18$-$30$ years) participants (all hearing thresholds $<= 30$ dB SPL, for $125$-$8000$ Hz), with Dutch/Flemish as their native language.
The subjects were recruited and presented with speech stimuli in Dutch, for a duration ranging between $90$ and $150$ minutes, divided into $6$ to $10$ recordings (i.e., an uninterrupted period in which a participant listens to a stimulus), totaling $168$ hours of EEG data. The EEG was collected using a BioSemi ActiveTwo system with 64 active Ag-AgCl electrodes, placed according to $10$-$20$ electrode system, with a sampling rate of $8192$Hz, where as the speech stimuli was sampled at $48$ kHz. More details about this data can be found in \cite{accou2023sparrkulee}.

\subsection{EEG Pre-processing}

The EEG pre-processing pipeline utilized in this study for both Natural and Dichotic datasets was based on the CNSP Workshop 2021 guidelines\footnote{https://cnspworkshop.net/resources.html} and implemented using the EEGLAB software~\cite{delorme2004eeglab}. 
First, a low-pass Butterworth filter with a cutoff frequency of $32$ Hz was applied to the EEG signal. This filter helps attenuate high-frequency noise and artifacts that are unrelated to the neural activity of interest. Next, a high-pass Butterworth filter with a cutoff frequency of $0.5$ Hz was employed to remove DC offsets or drifts in the signal. Following the filtering steps, the EEG data  were down-sampled to $64$ Hz. 

A channel rejection method based on the EEG data's variance was employed to identify and replace noisy channels. The channels with excessively high variance, indicating potential artifacts or poor signal quality, were considered unreliable and replaced using a spline interpolation technique. This interpolation is performed using the remaining channels to preserve the spatial information of the EEG data.

After channel rejection and interpolation, the EEG channels were re-referenced to a specific set of external channels known as mastoids. This re-referencing step helps minimize the effects of common noise sources and improves the interpretability of the EEG data by providing a reference point.

Lastly, to normalize the data and ensure consistency across channels and trials, a z-score transformation was applied. This process computes the z-score independently for each channel and each trial, by subtracting the mean followed by normalization with standard deviation. 

For the Natural Dataset - Large, we utilized the pre-processed data available at \url{https://doi.org/10.48804/K3VSND}. This data underwent standard pre-processing steps, consistent with those applied to the datasets mentioned earlier, resulting in a final sampling rate of $64$ Hz, as detailed in \cite{accou2023sparrkulee}.

\subsection{MM classification task for natural listening}

\begin{figure}[t!]
  \centering
  \includegraphics[width=\columnwidth]{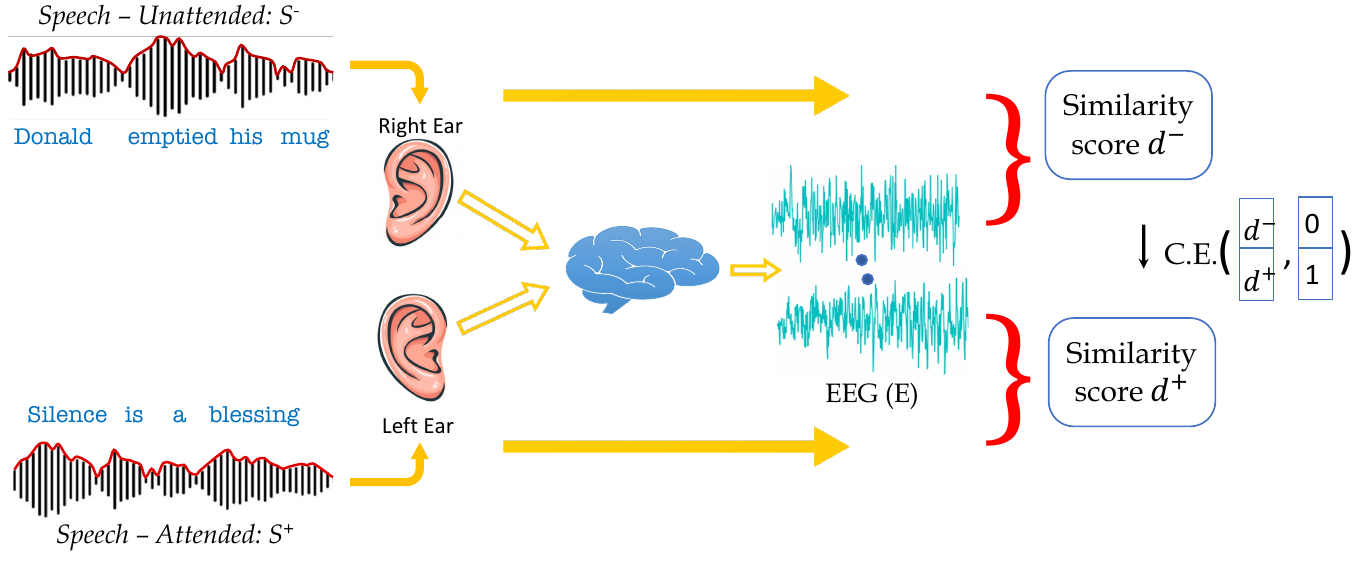}
  \centering
  \caption{
  \textbf{MM classification task on dichotic listening:} MM classification task is a binary classification paradigm associating the EEG and speech segments. The EEG segment ($\textbf{E}$) and corresponding attended stimulus sentence ($\textbf{S}+$) form the positive pair, while the same EEG and corresponding unattended sentence ($\textbf{S}-$) form the negative pair. The similarity score computation is achieved using the model depicted in Figure~\ref{fig:multi_nw}.  Here, C.E. denotes the cross-entropy loss. The dichotic listening condition is illustrated in this figure, where two distinct audio signals are simultaneously played to each ear of the listener. In contrast, during natural speech listening, the mismatched audio signal is chosen from a different speech region in the same session.}
  \label{fig:aadtask}
\end{figure}
\begin{figure*}[t!]
  \centering
  \includegraphics[width=\textwidth,trim=80 140 80 120,clip]{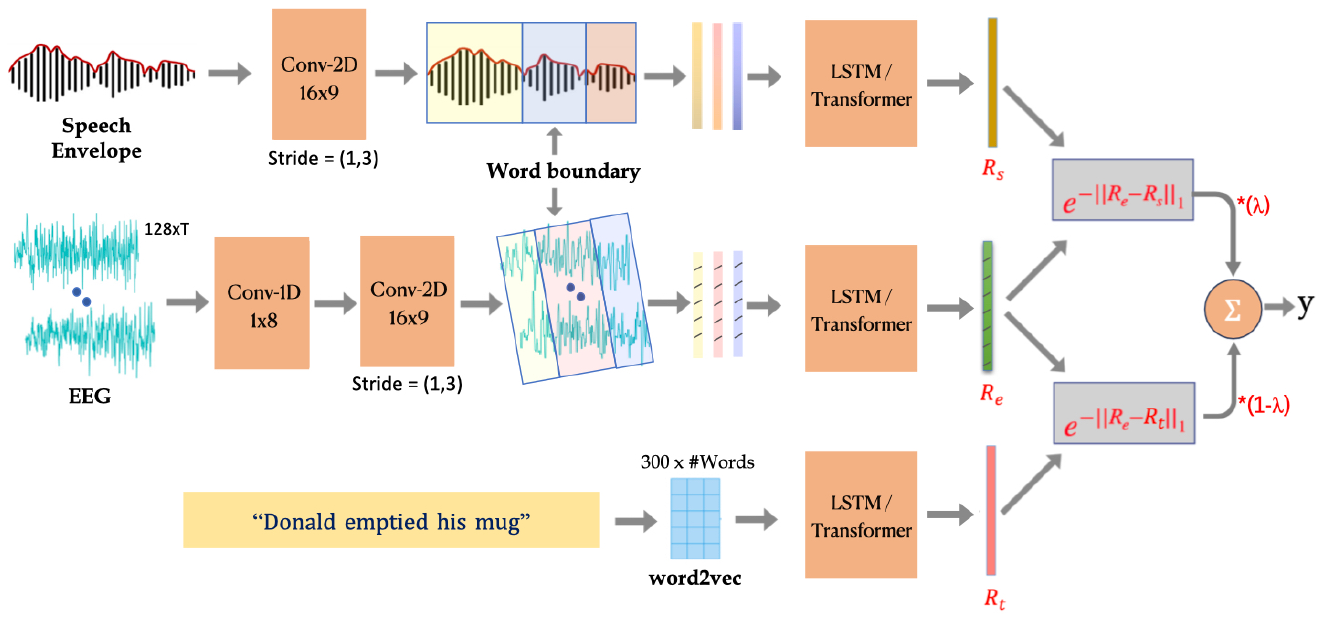}
  \vspace{-1cm}
  \caption{Proposed STEM3 using LSTM / Transformer architecture for MM task on text, speech and EEG  data.} 
  \label{fig:multi_nw}
  \vspace{-0.15in}
\end{figure*}
Our work uses the MM task as  the stimulus-response model to study the neural entrainment. The classification model is trained to maximize the similarity between the EEG and the ``attended'' speech segment while minimizing the similarity between the same EEG data and the ``un-attended'' audio. In this study, the speech segment is considered at a sentence-level. We also compare with prior works \cite{monesi2020lstm,jalilpour2021extracting}, which perform this task at the sentence level. In the natural listening condition, the time-aligned stimulus corresponding to the EEG response segment is the matched speech, while speech from another sentence in the same session is chosen as the mismatched speech. Selecting mismatched samples from the same trial makes the classification task challenging enough to encourage the model to learn the stimulus-response relationships. 
This sampling approach also avoids the chance of memorizing the speech features along with their labels.


\subsection{Auditory Attention Decoding as a MM task}
Figure~\ref{fig:aadtask} illustrates this paradigm in detail. The segment of speech that the subject is instructed to attend is considered as the match segment, while the segment of speech played in the other ear is considered as the mismatch segment. This task is more challenging than MM tasks on the natural speech dataset. The behavioral experiment that followed the listening experiment showed that the subjects were able to comprehend and understand the segment they attended to while having difficulty answering questions about the unattended speech. We employ a trial-selection based on behavioral scores (Figure~\ref{fig:trial-selection}), which is described in Sec.~\ref{sec:trial-selection}.


\subsection{Acoustic Feature - Speech Envelope}


The speech envelope represents the variations in the amplitude of the speech signal over time. It is obtained by extracting the magnitude of the signal's analytic representation, using the Hilbert transform. The temporal envelopes of sounds contain critical information for speech perception \cite{shannon1995speech,smith2002chimaeric}. It has been shown that auditory cortex can temporally track the acoustic envelopes \cite{poeppel2020speech}. The strength of cortical envelope tracking may serve as an indicator of the extent of speech perception (\cite{di2018cortical,etard2019neural,vanthornhout2018speech}).  Speech envelope provides valuable information about the overall shape and dynamics of the speech signal, including prosodic features, syllabic structure, and phonemic transitions. However, it should be noted that the speech envelope does not contain information regarding the semantics of the speech input.

\subsection{Semantic Feature - Word2vec text embedding}

The text transcription of the speech stimulus is provided in the dataset. This study used this text data to obtain features representing the semantics of the speech signal.  Semantic vectors for content words were derived using the word2vec algorithm \cite{mikolov2013efficient}. 

Word2Vec (w2v) is a popular word embedding technique proposed by Mikolov et al. \cite{mikolov2013efficient}. This algorithm generates a vector representation for each word. This study used pre-trained vectors trained on a subset of the Google News dataset (about 100 billion words). The model contains 300-dimensional vectors for 3 million words and phrases. The embeddings were obtained using a data-driven approach outlined in \cite{mikolov2013distributed}. These pre-computed word vectors are publicly available\footnote{https://code.google.com/archive/p/word2vec/}. The fundamental notion of w2v embedding is that words with similar semantics tend to be closer to each other in their vector space representation. In our work, the text embeddings were computed for each word in the sentence.  The w2v vectors of constituent words were concatenated together to obtain the sentence representation. This representation was fed as input to the models described in the following sections.

\subsection{Model architecture}
\label{sec:Model_architecture}


\subsubsection{Acoustic Encoding }
The speech signal representation $\mathcal{\mathbf{S}}$ is the speech envelope of dimension $1 \times T$, where $T$ denotes the duration of a speech sentence at $64$Hz. Similarly, the EEG data for the same sentence is denoted as $\mathcal{\mathbf{E}}$, which is of dimension $128 \times T$. 

The EEG sub-network (top row of Figure~\ref{fig:multi_nw})  consists of a series of convolutional layers and LSTM/Transformer layers. The convolutional layers implement $1$-D and $2$-D convolutions with $1 \times 8$ and $16 \times 9$ kernel sizes, respectively. Further, the $2$-D CNN layers also introduce a stride of $(1,3)$ to further down-sample the feature maps. The speech sub-network (top row in Figure~\ref{fig:multi_nw}) has a 2D convolutional layer of kernel size 16 and stride (1,3). For acoustic encoding, we utilize both the speech and EEG subnetworks in a parallel neural pipeline, as depicted in top half of Figure~\ref{fig:multi_nw}, without any weight sharing.

The word boundary information is incorporated at the equivalent sampling rate (both EEG and audio representations at $\frac{64}{3}$ Hz). The audio and EEG feature maps are average pooled at the word level using the word boundary information. As a result, for a given sentence, the EEG and speech branches generate vector representations sampled at the word level. The LSTM/Transformer layers model the inter-word context from these representations. This layer is included in both the stimulus (speech) and response (EEG) pathways.  The output of the LSTM/Transformer layer, of dimension $32$, is used as the embedding for the stimulus/response, denoted as $R_s$/$R_e$, respectively.  

We propose to optimize the Manhattan distance \cite{li2020combining} between the stimulus and response embeddings. The similarity score is computed as,
\begin{equation}\label{eq:eq1}
 \mathcal{\mathbf{Sim}}(\mathcal{\mathbf{E}}, \mathcal{\mathbf{S}}) = \exp (- || R_e - R_s ||_1)    
\end{equation}
\noindent{The similarity score for  matched pair ($\mathcal{\mathbf{E}}, \mathcal{\mathbf{S}^+}$) and mismatched pair ($\mathcal{\mathbf{E}}, \mathcal{\mathbf{S}^-}$) are computed. The model, with a dropout  of $0.2$, is trained using a binary cross-entropy loss, with [$\mathcal{\mathbf{Sim}}(\mathcal{\mathbf{E}}, \mathcal{\mathbf{S}^+})$, $\mathcal{\mathbf{Sim}}(\mathcal{\mathbf{E}}, \mathcal{\mathbf{S}^-})$] mapped to [$1$, $0$] targets.}

\subsubsection{Semantic Encoding }
To compare the encoding of text features with speech features, we use a text-EEG MM task. We used the text-EEG sub-network (bottom part of Figure~\ref{fig:multi_nw}) to learn the mapping between the w2v embeddings and the EEG response. The EEG sub-network resembles the acoustic encoding model, while the text sub-network consists of two-layers of LSTM/Transformer architecture with w2v embeddings at word-level as input features. The last hidden state of text sub-network is used as text embedding, denoted as $R_t$. The similarity scoring component is identical to the acoustic encoding model (Eq.~\ref{eq:eq1}).

\subsubsection{Joint Encoding of Acoustics and Semantics }
In addition to exploring the effect of acoustic and semantic features individually, we perform an investigation of the joint training of the STEM3 with both the modalities. The sub-networks are combined by defining a joint-loss function over the two modalities (detailed in Sec.~\ref{sec:multimodal-similarity-measure}). 
For a given match segment, the EEG response corresponds to the attended audio/text features, while in the mismatched example, the audio/text features are chosen from a different region (natural speech) or from the stimuli stream played in the other ear (dichotic listening). 

To effectively process the Natural Dataset - Large and mitigate the issue of underfitting, the LSTM layer is replaced with a transformer encoder. This modification enhances the model's capacity to capture complex patterns and dependencies within the larger SparrKULee data. The LSTM architecture is used on Natural Dataset - small and the Dichotic dataset.

\subsection{Trial selection in Dichotic listening}\label{sec:trial-selection}
\begin{figure}[t!]
  \centering
  \includegraphics[width=0.75\linewidth]{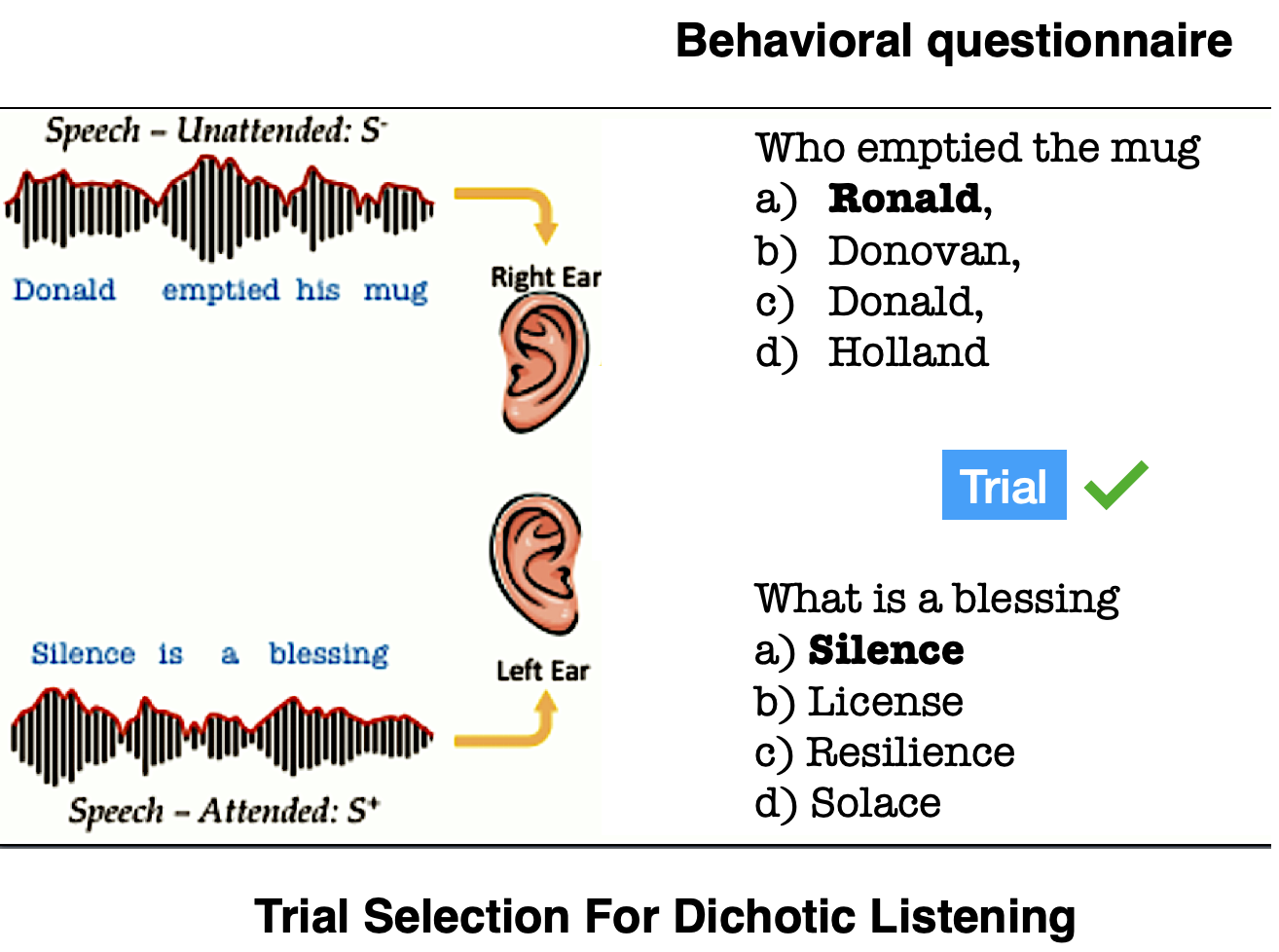}
  \caption{\textbf{Trial selection for dichotic listening:} At the end of the dichotic listening, a questionnaire is presented to the listeners. The trials, where the participants correctly answer the attended side (according to the prior instruction) with  a score   $>60$\%, while incorrectly responding to the unattended side (score of $<40$\%), are chosen.  }
  \label{fig:trial-selection}
\vspace{-0.15in}
\end{figure}

As part of the dichotic listening task, participants were required to answer multiple-choice questions after each trial to assess their comprehension and, consequently, their attentiveness to the played stories \cite{broderick2018electrophysiological}. They were asked $10$ questions each from  both the stories. The scores obtained were normalized to a range of $0$ to $1$. For all the experiments that we report in this work,  we include   trials in which the participants achieved a comprehension score of $60$\% or more for the attended audio, while achieving a score of $40$\% or less for the un-attended audio. This choice of scores for attended / un-attended streams ensured that the ``matched'' / ``mismatched'' EEG-audio/text pairs  had  a clear distinction. The selection is elicited in Figure~\ref{fig:trial-selection}. While this selection led to a removal of $35$\% of the total trials, the process ensured a reliable label for the training and evaluation.  

\subsubsection{Training and Evaluation Setup } 
\label{sec:train_setup_c4}
The experiments are run with a batch size of $32$. The models are trained using Adam optimizer \cite{kingma2014adam} with a learning rate of $0.001$ and weight decay parameter of $0.0001$. The models are learned with a binary cross-entropy loss.


 Overall,  the Natural dataset - Small contains $20$ stimulus trials, while the dichotic listening dataset contains $30$ trials. A subset of trials from all subjects constituted the training set, while the remaining unseen trials of all subjects served as the test data. In each validation fold, we randomly chose $3$ trials  to form the test data.  This resulted in $6$-fold cross-validation for the natural speech dataset and $10$-fold cross-validation for the cocktail party dataset. The average classification accuracy across validation folds is reported in the results section. In Natural dataset - Large, out of $85$ subjects, $14$ subjects' data were kept  for evaluation and the remaining were used in training and validation in $90:10$ ratio.


To assess statistical significance, we employed the Wilcoxon signed-rank test \cite{demvsar2006statistical}. This is performed in a paired setting, where the predicted probabilities from the two models are used in the comparison. 
We have used an $\alpha$ level of $.01$ for determining the statistical significance. For the unpaired t-test, we have also performed the Wilcoxon rank-sum test.


\subsection{Multimodal Loss Function and Joint training}\label{sec:multimodal-similarity-measure}

We explore three choices to combine the modalities,
\begin{align}
Sim_1(\mathcal{\mathbf{E}}, \mathcal{\mathbf{S}}, \mathcal{\mathbf{T}}) 
&= \lambda \cdot \exp \left( - \| R_s - R_e \|_1 \right) \nonumber \\
&\quad + (1-\lambda) \cdot \exp \left( - \| R_t - R_e \|_1 \right) \label{eq:sim1} \\
Sim_2(\mathcal{\mathbf{E}}, \mathcal{\mathbf{S}}, \mathcal{\mathbf{T}}) 
&= \exp \left( - \| R_s - R_e \|_1 \right)^\lambda \nonumber \\
&\quad + \exp \left( - \| R_t - R_e \|_1 \right)^{1-\lambda} \label{eq:sim2} \\
Sim_3(\mathcal{\mathbf{E}}, \mathcal{\mathbf{S}}, \mathcal{\mathbf{T}}) 
&= \exp \left( - \left\| \lambda \cdot R_s + (1-\lambda) \cdot R_t - R_e \right\|_1 \right) \label{eq:sim3}
\end{align}
Here, $\lambda = 0$ implies that only the semantic network (text-EEG) is considered,  $\lambda =1$ implies that only the acoustic network (speech-EEG) is considered while $\lambda =\{0.2,0.4,0.5,0.6,0.8\}$ implies joint modeling (EEG-speech+text). 
During training, the speech, text and EEG data are loaded and the value of ${\lambda}$ is randomly selected from $\{0.0, 0.5, 1.0\}$. 
In this way, a single model is trained with all three combinations, of text alone ($\lambda = 0$), speech alone ($\lambda = 1$) and speech-text multimodal ($\lambda = 0.5$).
During inference, the model can be input with EEG-text samples ($\lambda = 0$), EEG-speech samples ($\lambda = 1$) as well EEG-speech+text ($\lambda = 0.5$). This flexibility helps us understand the role of acoustics and semantics from the same model.
\begin{figure}[t!]
  \centering
  \includegraphics[width=1\linewidth]{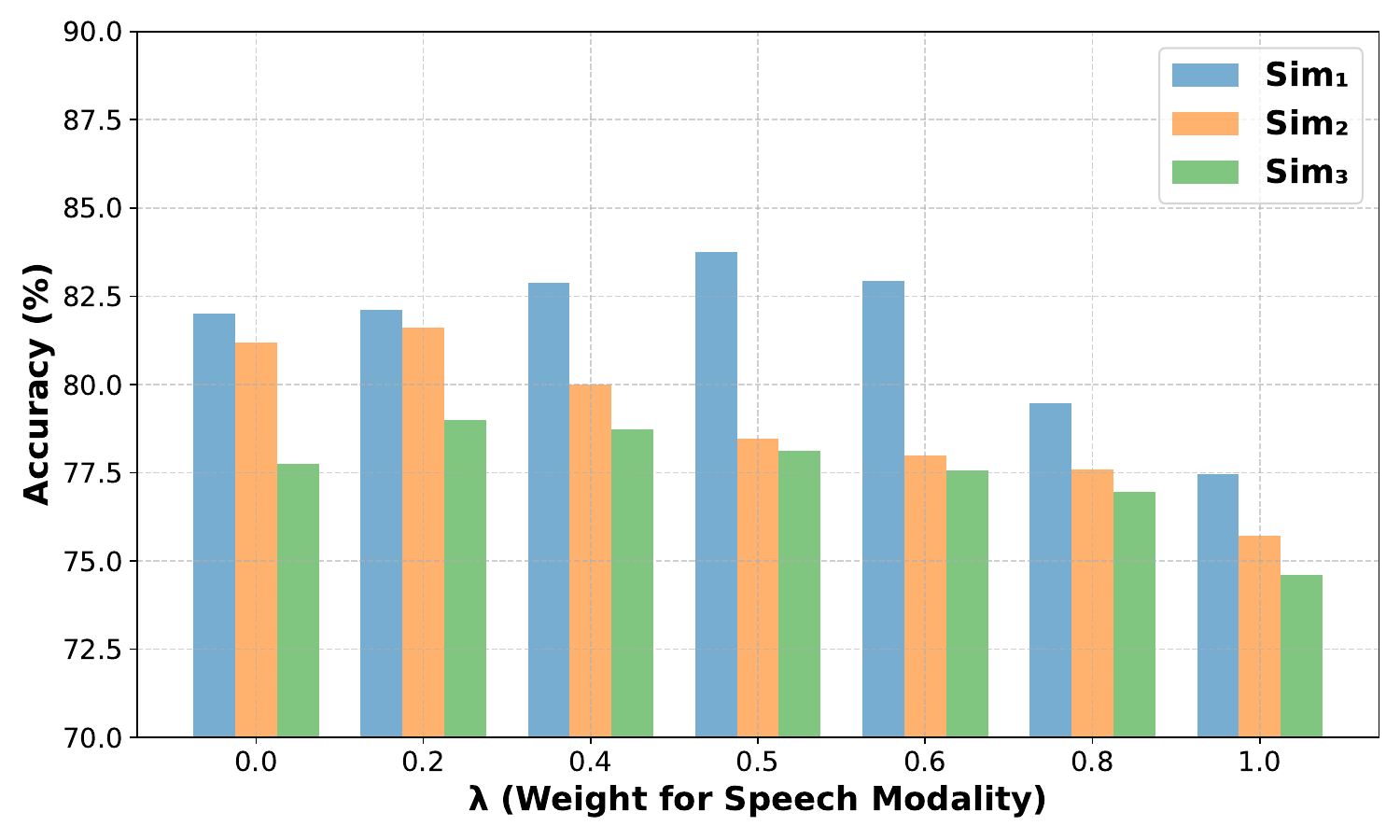}
  \caption{\textbf{Average MM classification accuracy(\%) with different similarity measures on dichotic listening:} Here, $\lambda$ denotes the weight applied to the acoustic (speech-EEG) component in the similarity function. }
  \label{fig:Similaritywise_accuracies}
\vspace{-0.15in}
\end{figure}


\section{Results}
\subsection{Choice of loss function}
For dichotic listening condition, we experiment with different $\lambda$ values during inference. 
As shown in Figure~\ref{fig:Similaritywise_accuracies}, it is observed that
 semantic encoding in EEG achieves a higher MM performance than the acoustic encoding network. Further, the similarity measure \(Sim_1(\mathcal{\mathbf{E}}, \mathcal{\mathbf{S}}, \mathcal{\mathbf{T}})\) proved to be best performing 
 similarity measure. The best MM classification results are achieved with $\lambda\ = 0.5$.  All the subsequent experiments  use this configuration with \(Sim_1(\mathcal{\mathbf{E}}, \mathcal{\mathbf{S}}, \mathcal{\mathbf{T}})\)  similarity measure.


\subsection{Impact of Acoustic and Semantic Cues}

\begin{table*}[t]
\centering
\caption{MM classification accuracy (\%) of speech stimulus and EEG responses across different datasets and models. Average results for various feature modalities are reported, including baseline model performance.}
\label{tab:diff_feat_sd}
\begin{tabular*}{\textwidth}{@{\extracolsep{\fill}}lllll}
\hline
 &  & \textbf{Speech Envelope} & \textbf{Text Word2vec} & \textbf{Multimodal} \\
\textbf{Dataset} & \textbf{Model} & \textbf{(Acoustic) } & \textbf{(Semantic) } & \textbf{(Acoustic + Semantic) } \\
 &  & (\(\lambda=1\)) & (\(\lambda=0\)) & (\(\lambda=0.5\)) \\
\hline
 & Baseline \cite{monesi2020lstm} & 65.23 & -- & --  \\
Natural Dataset - Small & Borsdorf \textit{et al.}~\cite{borsdorf2023multi}$^{\rm a}$ & 68.96 & -- & -- \\
 & Wang \textit{et al.}~\cite{wang2024self}$^{\rm b}$ & -- & -- & 36.33 \\
 & \textbf{STEM3} & \textbf{93.63} & 93.24 & \textbf{93.38} \\
\hline
 & Baseline \cite{jalilpour2021extracting} & 62.00 & -- & -- \\
Dichotic Dataset & Borsdorf \textit{et al.}~\cite{borsdorf2023multi}$^{\rm a}$ & 49.98 & -- & -- \\
 & Wang \textit{et al.}~\cite{wang2024self}$^{\rm b}$ & -- & -- & 56.01 \\
 & \textbf{STEM3} & \textbf{77.45} & 82.00 & \textbf{83.76} \\
\hline
 & Baseline \cite{monesi2020lstm} & 50.05 & -- & -- \\
Natural Dataset - Large$^{\rm c}$ & Borsdorf \textit{et al.}~\cite{borsdorf2023multi}$^{\rm a}$ & 74.77 & -- & -- \\
 & Wang \textit{et al.}~\cite{wang2024self}$^{\rm b}$ & -- & -- & 50.33 \\
 & \textbf{STEM3}$^{\rm d}$ & \textbf{88.70} & 74.37 & \textbf{87.91} \\
\hline
\end{tabular*}

\vspace{1mm}
\begin{minipage}{\textwidth}
\footnotesize
$^{\rm a}$ Envelope with MHA+DC model was selected to maintain uniformity in comparison. \\
$^{\rm b}$ Envelope + GPT model was selected for consistency; “$\lambda=0.5$” does not apply here. \\
$^{\rm c}$ All results with Natural Dataset - Large were evaluated as described in Section~\ref{sec:Dataset}. \\
$^{\rm d}$ Here, LSTM was replaced with a transformer as discussed in Section~\ref{sec:Model_architecture}.
\end{minipage}
\end{table*}

Table~\ref{tab:diff_feat_sd}  presents the results of the MM classification accuracy between speech stimuli and their corresponding EEG responses, focusing on different listening conditions.
Our \textbf{STEM3} was evaluated alongside the baseline model \cite{monesi2020lstm} \cite{jalilpour2021extracting} and two other models \cite{borsdorf2023multi} \cite{wang2024self} across three datasets.
\textbf{STEM3} demonstrated an average relative improvement of $79\%$ on the Natural Dataset – Small, $50.73\%$ on the Dichotic dataset, and $57.22\%$ on the Natural Dataset – Large, with all improvements being statistically significant(Wilcoxon signed-rank test, $\textit{p}<<0.01$) as shown in Table~\ref{tab:diff_feat_sd}.

Compared to the MM classification of natural speech under normal listening conditions, the task of auditory attention decoding (AAD) in dichotic listening is more challenging. 

\begin{figure*}[t!]
  \centering
  \includegraphics[width=\textwidth]{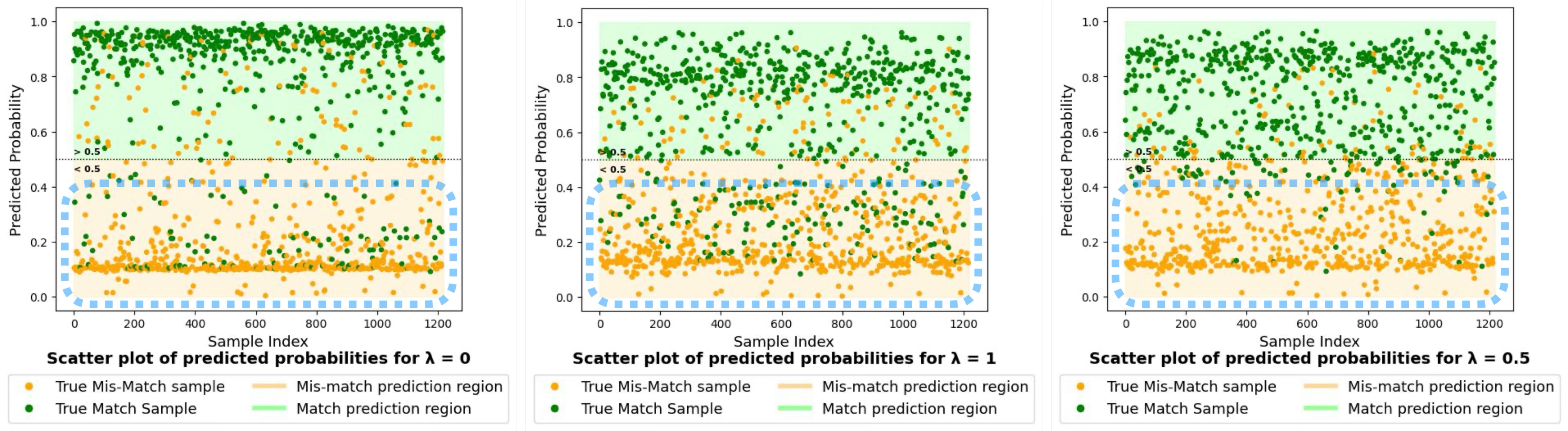}
  \caption{\textbf{Scatter plots for predicted probabilities in dichotic listening task :} From left to right, the scatter plots for $\lambda =\left\{0, 1, 0.5\right\}$ are shown. A region in the mismatched prediction ($0-0.4$) is highlighted to indicate how the semantic (text-EEG) MM task is visibly better than the acoustic (speech-EEG) MM task, in terms of mis classified examples. The joint model reduced the mis-classifications in the highlighted region. }
  \label{fig:Scatter Plot}
  \vspace{-0.1in}
\end{figure*}

For natural speech conditions, both semantic and acoustic features result in similar accuracy (Wilcoxon signed-rank test, $\textit{p} > 0.01$).
For dichotic listening, word2vec features result in significant improvement in accuracy compared to speech envelope features (Wilcoxon signed-rank test, $\textit{p}<<0.01$). The scatter plot for prediction probabilites for dichotic listening condition and for a fold in validation is shown in Figure~\ref{fig:Scatter Plot}.

\subsubsection{Importance of word boundary information }

Word-level segmentation holds significant importance in speech perception during normal listening conditions \cite{akshara2023IS}. To investigate the role of word segmentation in auditory attention decoding, we conducted a comparison of MM classification task performance using the proposed model with and without word boundary information in the dichotic listening condition. The word boundary segmentation is used only for the EEG and acoustic features, while the word2vec features are inherently sampled at word-level.

As seen in Table~\ref{tab:wb_aad}, word level segmentation plays an important role in auditory attention decoding. Word boundary information has impact on both semantic features (Wilcoxon rank-sum test, $\textit{p}<<0.01$) and  envelope features (Wilcoxon rank-sum test, $\textit{p}<<0.01$). Word boundary information also shows a significant impact on the joint speech-text model.

\begin{table}[t]
\centering
\caption{Role of word boundary information in auditory attention detection. The table shows the MM classification accuracy (\%) of speech stimulus and its EEG responses for the dichotic listening condition. The model was trained with and without word boundary information for comparison.}
\label{tab:wb_aad}
\begin{tabular}{lcc}
\hline
\textbf{Stimulus Feature} & \textbf{Without} & \textbf{With} \\
 & \textbf{Word Boundary} & \textbf{ Word Boundary} \\
\hline
Env (\(\lambda=1\)) & 67.09 & 77.45 \\
w2v (\(\lambda=0\)) & 77.35 & 82.00 \\
Both (\(\lambda=0.5\)) & 76.69 & \textbf{83.76} \\
\hline
\end{tabular}
\vspace{-0.1in}
\end{table}

\subsubsection{Importance of accurate word boundaries }
We conducted another ablation study to understand the impact of accurate word boundary information. In this setting, the model is fed with random word boundaries with varying granularity. Each sentence is segmented to a fixed number of regions, and their boundaries are chosen at random.  The results are reported in Table~\ref{tab:randomWB_aad}.
The performance does not significantly improve for speech-EEG (Wilcoxon rank-sum test, $\textit{p}=0.030$), text-EEG (Wilcoxon rank-sum test, $\textit{p}=0.406$), or with multimodality (Wilcoxon rank-sum test, $\textit{p}=0.090$), even as the number of random word boundaries increases. However, the use of accurate word-boundaries is significantly better than those with random word boundaries ($\textit{p} << 0.01$).


\begin{table}[t]
\centering
\caption{Average MM classification accuracy (\%) for auditory attention detection with random word boundaries and the true word boundary.}
\label{tab:randomWB_aad}
\begin{tabular}{lccccc}
\hline
\textbf{Stimulus} & \textbf{2} & \textbf{3} & \textbf{4} & \textbf{5} & \textbf{True} \\
\textbf{Feature} &  &  &  &  & \textbf{Word Boundary} \\
\hline
Env (\(\lambda=1\))       & 66.13 & 66.16 & 67.39 & 70.89 & 77.45 \\
w2v (\(\lambda=0\))       & 78.95 & 76.29 & 78.58 & 78.19 & 82.00 \\
Both (\(\lambda=0.5\))    & 79.11 & 79.20 & 79.55 & 79.41 & \textbf{83.76} \\
\hline
\end{tabular}
\end{table}

\subsubsection{Region wise Analysis }

\begin{figure}[t!]
  \centering
  \includegraphics[width=0.9\linewidth]{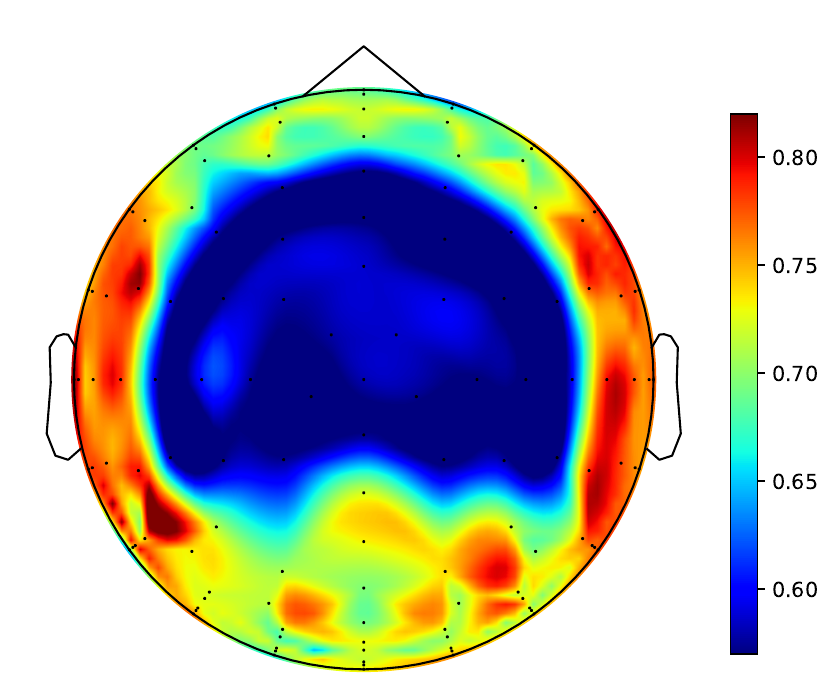}
\caption{\textbf{MM classification accuracy across brain regions:} Heat map illustrating the model’s classification accuracy segmented by brain regions, based on 128-channel BioSemi EEG system for \(\lambda = 1\). The results highlight notably higher accuracy in the temporal region compared to others.}
  \label{fig:Brain_regions}
  \vspace{-0.1in}
\end{figure}

\begin{table}[t]
\centering
\caption{MM classification accuracy (\%) of speech stimulus and EEG responses broken down by brain region, for \(\lambda\) values of \{0, 0.5, 1\}, in the dichotic listening task.}
\label{tab:region_wise}
\begin{tabular}{lccc}
\hline
 & \textbf{Speech} & \textbf{Text} & \textbf{Multimodal} \\
\textbf{Brain} & \textbf{Envelope} & \textbf{Word2vec} & \textbf{(Acoust. +} \\
\textbf{Region} & \textbf{(Acoust.)} & \textbf{(Sem.)} & \textbf{ Sem.)} \\
 & \textbf{(\(\lambda=1\))} & \textbf{(\(\lambda=0\))} & \textbf{(\(\lambda=0.5\))} \\
\hline
Frontal   & 69.70 & 74.85 & 75.65 \\
Central   & 57.39 & 62.30 & 62.47 \\
Parietal  & 71.09 & 78.03 & 78.57 \\
Temporal  & 77.00 & 79.33 & \textbf{81.14} \\
Occipital & 73.33 & 79.00 & 79.98 \\
\hline
\end{tabular}
\end{table}

We performed an experiment to understand which regions of the brain were relatively more involved and influential in performing the attention task in a dichotic listening setting. While spatial localization in EEG is not accurate, the analysis may indicate potential regions which carry more information for the MM task. 

The electrodes of the $128$-electrode EEG setup are segregated into five regions namely frontal, central, parietal, temporal and occipital. We use the $10$-$20$ map to perform this segregation. 
Only the  subset of electrodes belonging to a region are used in training and cross-validation. As shown in Table~\ref{tab:region_wise}, the temporal region performs significantly better in MM classification in comparison with other regions (Wilcoxon rank-sum test, $\textit{p}<<0.01$). The performance heat-map (Figure~\ref{fig:Brain_regions}), measured for the electrodes belonging to various brain regions, showed a significant preference to the temporal region.

\subsubsection{Right Ear Advantage }
\begin{table}[t]
\centering
\caption{MM classification accuracy (\%) of speech stimulus and EEG responses based on subjects who attended to speech stimuli played on the left (S1) or right (S2) ear.}
\label{tab:Stimulus_wise}
\begin{tabular}{lccc}
\hline
 & \textbf{Speech} & \textbf{Text} & \textbf{Multimodal} \\
\textbf{Brain} & \textbf{Envelope} & \textbf{Word2vec} & \textbf{(Acoust. +} \\
\textbf{Region} & \textbf{(Acoust.)} & \textbf{(Sem.)} & \textbf{ Sem.)} \\
 & \textbf{(\(\lambda=1\))} & \textbf{(\(\lambda=0\))} & \textbf{(\(\lambda=0.5\))} \\
\hline
S1 (left ear)  & 74.14 & 79.12 & 83.52 \\
S2 (right ear) & 79.55 & 84.75 & 87.37 \\
\hline
\end{tabular}
\end{table}

We conducted a set of ablation experiments aimed at identifying which stimulus group (attention to left versus right ear) elicited the strongest performance in the MM classification task for dichotic listening task. As mentioned in Section~\ref{sec:trial-selection}, the listeners are instructed to attend to stories played in left or right ear. 
Thus, we split the data into two parts, based on the story the participant was instructed to attend, and chose the trials which positively reflected this instruction in the behavioral data.
An initial observation  was that the trial selection (discussed in Section~\ref{sec:trial-selection}) retained more trials ($67.48$ \%) when the subjects were asked to attend to the right ear than when they were asked to attend to the left ear ($65.27$\%). 
The MM task results from the two subject splits are shown in Table~\ref{tab:Stimulus_wise}, after balancing the training data of each group to the same number of samples.

We observed that the subject group instructed to listen to the right ear (S2) had significantly improved MM task performance  than  the subject group (S1) instructed to attend to the left ear  (Wilcoxon rank-sum test \(\textit{p}<<0.01\)). This improvement was consistently seen on all modalities (speech, text and multimodal).

\section{Discussion}
\subsection{Word Boundary Information}
With a set of experiments, we have attempted to validate the hypothesis that speech perception in the brain is segmented at the word level. The incorporation of word boundary information yields statistically significant improvements compared to the baseline model, demonstrating the impact of this information in the neural tracking of speech. 
Both normal and dichotic listening tasks benefit from the use of accurate word information. 
Further, random choice of word-boundaries does not have any significant impact on the task. 
This result supports the notion that the human brain prioritizes comprehending the content over phonological aspects of the stimulus during challenging listening conditions, and this comprehension likely occurs at the word level. 

\subsection{Natural Speech: Speech Envelope vs Semantics}
Table~\ref{tab:diff_feat_sd} displays the outcomes of the MM classification task conducted using two distinct stimulus features: speech envelope and word2vec features. Our findings suggest that modelling EEG responses and stimuli yield superior performance during natural speech perception. Furthermore, we observed that the relative importance of textual and acoustic content is approximately similar (Wilcoxon signed-rank test, $\textit{p} > 0.01$) in the natural listening condition (small). Therefore, it can be concluded that both acoustic and semantic features hold comparable importance in speech perception during normal listening conditions. In natural listening condition (large), we find the acoustic features to be significantly better than textual features (Wilcoxon signed-rank test, $\textit{p} << 0.01$).

\subsection{Dichotic Listening: Speech Envelope vs Semantics}


During dichotic listening, the human brain exhibits a prioritization of perceiving the semantics over the acoustic characteristics of the attended speech (Wilcoxon signed-rank test, $\textit{p} << 0.01$). This indicates that semantic features are better entrained in the EEG response recorded during the dichotic listening experiment. The multimodal network achieves an average accuracy of $83.76$\%, compared to $77.45$\% with the speech feature-based network. This significant performance improvement (Wilcoxon signed-rank test, $\textit{p}<<0.01$) indicates that the EEG signal jointly encodes the semantic and acoustic content of the stimulus. The findings also suggest that semantics may hold higher significance than acoustic traits for speech perception under complex listening conditions. Furthermore, this MM performance signifies that the proposed multimodal model is able to identify the intended channel of attention in $83$\% of the segments, a capability that can potentially influence in cognitive design of hearing aids.

\subsection{Region-wise analysis}
The region-wise analysis, shown in Table \ref{tab:region_wise}, revealed that the temporal region exhibited the highest performance in the mismatch (MM) classification task, followed by the parietal, occipital, frontal, and central regions. The performance trend is also observed to be consistent for semantic and acoustic features. The observed pattern (Figure \ref{fig:Brain_regions}) suggests that temporal brain regions may play more critical roles in processing dichotic stimuli, particularly in tasks involving attention and speech processing. These findings are also consistent with those reported by Hausfeld et al.~\cite{hausfeld2024fmri} and Ross et al.~\cite{ross2010temporal}.

\subsection{Right ear advantage in dichotic listening}
Additionally, the results presented in Table~\ref{tab:Stimulus_wise} show that subjects exhibited significantly higher attention to the stimulus presented in the right ear (S2) compared to the left ear (S1) (Wilcoxon rank-sum test \(\textit{p}<<0.01\)). This supports the phenomenon of right ear dominance \cite{geffen1978development}, which is often considered as neurophysiological bias especially for speech stimulus in dichotic listening tasks, where individuals typically pay greater attention to stimuli delivered to the right ear. We find these results to align with observations of Tanaka et al.~\cite{tanaka2021neurophysiological} on the asymmetry of hemispheric organization and with Kazimierczak et al.~\cite{kazimierczak2022combined} on auditory laterality in brain.
\subsection{Limitations and Future Directions}
The lack of sharply defined functional boundaries in the brain and the inherent complexity of brain activity limit the precision of analyses based on non-invasive EEG signals. While the proposed architectures for normal and dichotic listening offer valuable insights into the semantic and acoustic components of attention, the current experimental setup does not fully reflect real-world conditions. Factors such as multiple spatially distributed sound sources, listener orientation, and the influence of visual and other sensory cues—each of which might affect auditory attention—are not accounted for.
Future work will focus on extending the paradigm to incorporate more ecologically valid scenarios by integrating these real-world variables into stimulus design. 

\section{Conclusion}
In this study, we compared Stimulus-EEG match-mismatch models under both natural speech and dichotic listening conditions, aiming to assess the relative contribution of acoustic and semantic features in different listening scenarios. 
Our findings reveal that the significance of textual and acoustic content is similar during natural speech listening. However, in the dichotic speech listening task, EEG signals demonstrated a higher MM performance for textual data compared to the audio signal. This suggests a more pronounced encoding of higher-level semantic information over acoustic envelope information in dichotic listening condition.
The study also investigated the role of word boundary information in the MM task performance.
It is seen that the word boundary information plays a pivotal role in speech perception during the dichotic listening task. Further, the study added further evidence  to right ear advantage observed in dichotic listening  through the MM classification experiments. 

\bibliographystyle{ieeetran}
\bibliography{references}

\end{document}